\documentclass{ceab}  

\usepackage{epsfig}    
\usepackage{graphicx}   

\usepackage{ceabbib}     

\usepackage[T1]{fontenc}

\begin{document}

\title{North-south differences of solar cycles}

\author{J. Murak\"ozy and A. Ludm\'any
\vspace{2mm}\\
\it Heliophysical Observatory, Hungarian Academy of Sciences, \\
\it H-4010 Debrecen, P.O.Box 30. Hungary}

\maketitle

\begin{abstract}
A long-term variation is detected in the north-south differences of the solar activity, the phase difference of hemispheric cycles alternates by four Schwabe cycles. This variability is demonstrated by two different methods. It can be the related to the Gleissberg cycle. The variation of the sunspot group tilts is similar to the hemispheric phase lags on a shorter timescale. Some possible relations are mentioned.

\end{abstract}

\keywords{solar cycle, north-south asymmetry}

\def\gore{N/S asymmetry} 

\section{Introduction}

Asymmetry is an important feature of astrophysical dynamos and its properties may be important in understanding the specific processes. In the case of solar dynamo several asymmetric properties are well discussed, they may have important roles to maintain the dynamo process in spite of the Cowling theorem. The mirror asymmetry of the rotating vortices due to the Coriolis force in the northern and southern hemispheres (Murak\"ozy and Ludm\'any, 2008) are important ingredients of the alpha effect. Deviations from the axial symmetry manifest themselves in a multipole structure by producing active longitudes (Berdyugina et al., 2003). These are also important in the active stars (Ol\'ah et al., 2009). 

The present work focuses on the solar north-south asymmetry with special emphasis on its variations. The early works in this field are based on sunspot data (Newton and Milsom, 1955) and reported no regular variations. Later works resulted in a diversity of temporal variations. Of course, the obtained different periods don't exclude each other but further studies are necessary to clarify the most important types of variations. We will return to these results in the discussions.

\section{Observational data}

The observational bases of the work are the Greenwich Photoheliographic Results (GPR) for cycles 12-20, the Kislovodsk sunspot data for cycle 21 and the Debrecen Photoheliographic Data (DPD, Gy\H ori et al. 2009) for cycles 22-23. The combined dataset from different sources may be inhomogeneous from several aspects but here we focus on the north-south asymmetry and it is the same for images of any sources. The institutional biases are the same for the two hemispheres and by applying any correction factors for both hemispheres the rate of asymmetry would not be distorted. This assumption has also been checked by different homogenization procedures and we will demonstrate the source-independence of the results.

\section{Hemispheric phase differences}

The first method to study the hemispheric phase lags is based on the cycle profiles. The center of mass has been computed for each hemispheric cycle profile so that the height of this point represents the strength of the cycle and its temporal position shows the date to which the cycle as a whole can be bound. This approach doesn't consider the highly irregular shapes of the cycle profiles which could mislead the comparisons, in this method the bulk of the hemispheric cycle, i.e. the torus as a whole is considered regardless of the short-term fluctuations. Fig.1. shows the cycles 12-23 for the northern and southern hemisphere with the centers of mass.

    \begin{figure}
  \begin{center}
   \epsfig{file=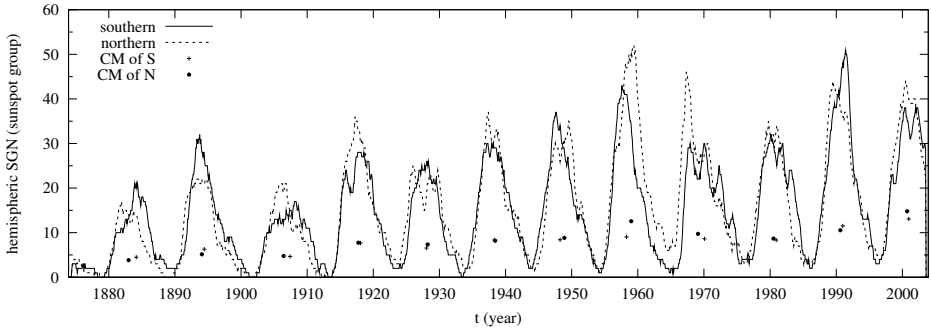,angle=-90,width=12.7cm}
  \end{center}
  \caption{Cycle profiles in the northern and southern hemispheres along with 
the centers of mass of the cycle profiles.}
\end{figure} 

Figure 2. shows the temporal differences between the centers of mass of both hemispheres. The vertical axis shows the difference between the dates of northern and southern centers of mass so negative values mean that the northern cycle precedes. It is conspicuous that the phase difference gradually changes from northern precedence to southern between cycles 12-19 and then, by an abrupt reversal, the southern hemisphere takes again the leading role and the trend starts again. The figure shows two panels, the first one is based on the original data of the three data sources whereas the second one is calibrated in the following way. A correction factor has been obtained for the Kislovodsk data by using the Greenwich-Kislovodsk data ratios in their overlapping period and the procedure was applied for the Debrecen data based on the Kislovodsk-Debrecen overlap. The two figures are practically identical, the largest difference between two phase-lag data is two days, unperceivable in the plots. The reason of the differences is that the multiplication of all data in a cycle with a correction factor may shift the date of the minimum between the neighboring cycles and of course this also shifts the date of the center of weight. Anyhow, one can conclude that the insitutional biases do not distort the rates of precedence between the two hemispheres.

    \begin{figure}
  \begin{center}
   \epsfig{file=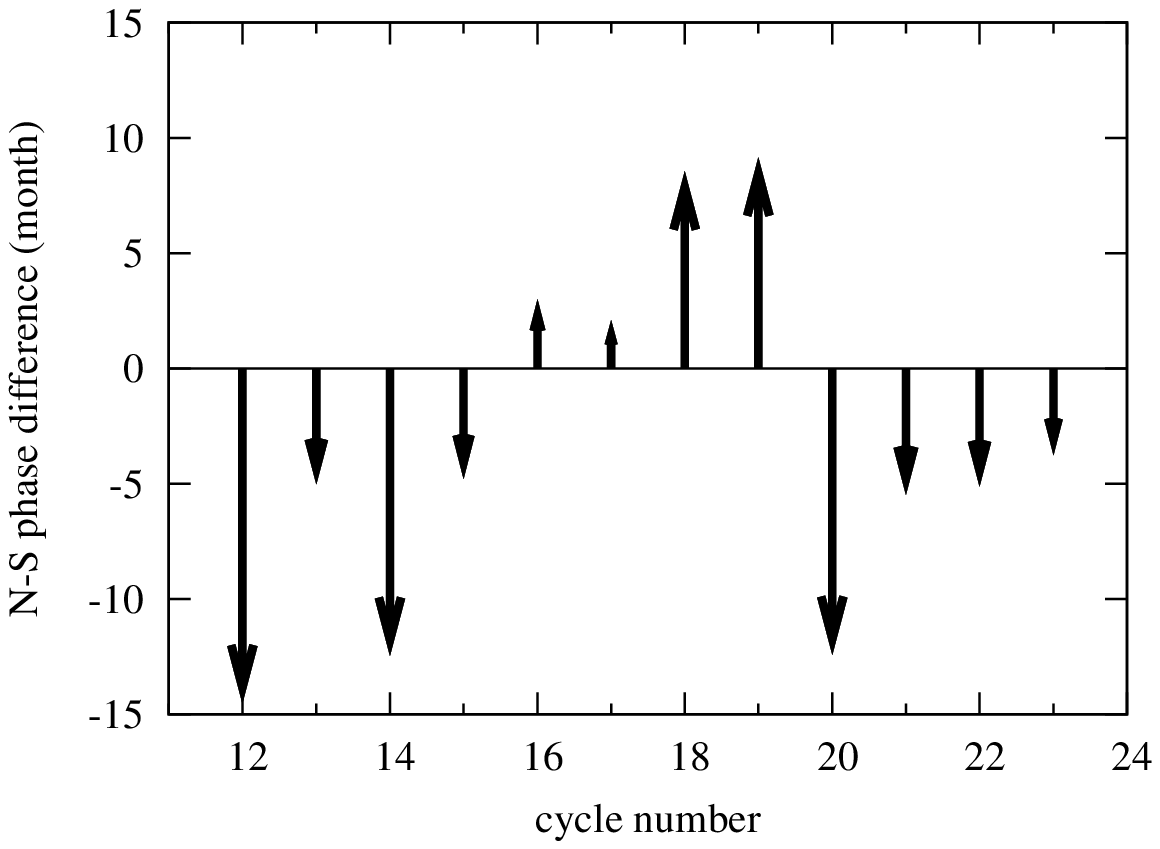,angle=-90,width=10cm}
   \epsfig{file=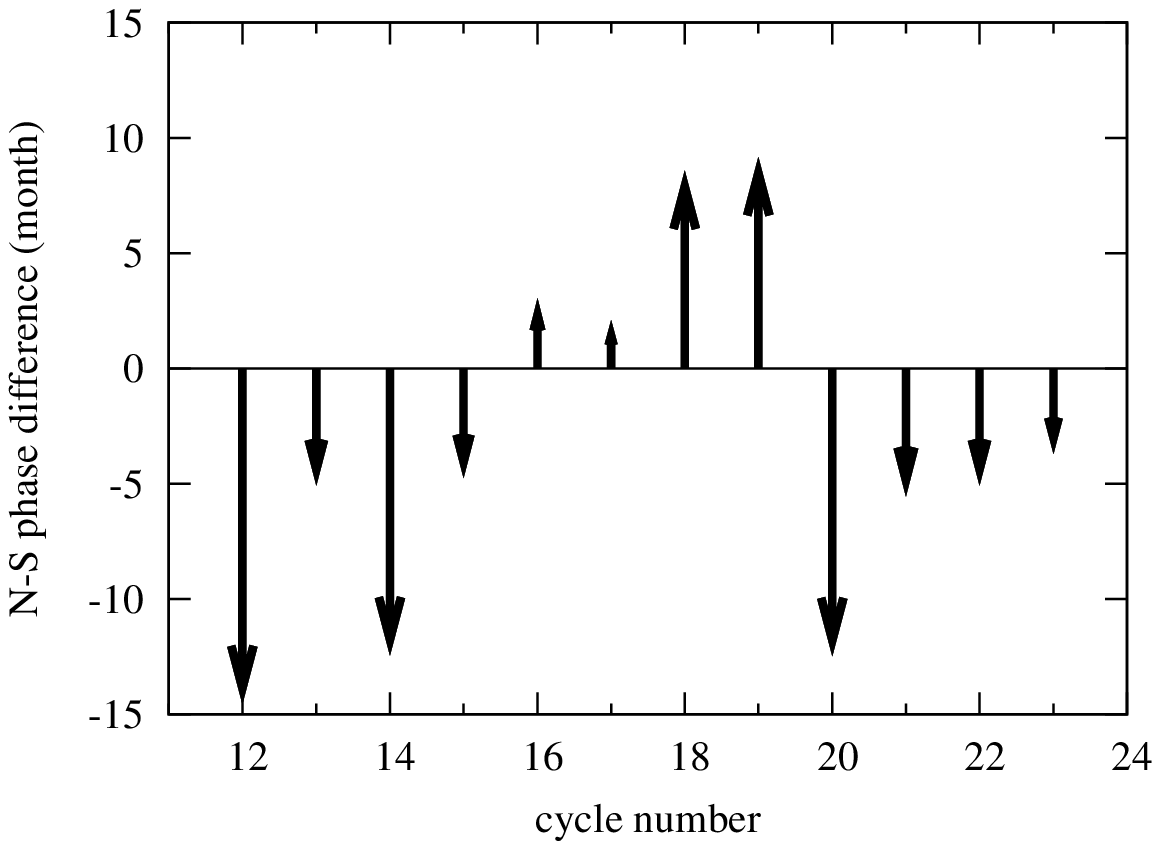,angle=-90,width=10cm}
  \end{center}
  \caption{Temporal differences between the centers of mass in the northern and southern cycles. Negative values: the northern cycle is leading. Upper panel: original sunspot data from the three applied catalogues, lower panel: calibrated data (see text).}
\end{figure} 

The phase lags can also be studied by a different method. The normalized north-south asymmetry index form sunspot group numbers (SGN) is: 

\begin{equation}
      NA = \frac{SGN_{N} - SGN_{S}}{SGN_{N} + SGN_{S}}
\,,
   \end{equation}

It is plotted for the period under study in Fig.3. The ascending and descending phases of the cycles are indicated by grey and white stripes. One could hardly recognize any regularity in the curve but if one plots the asymmetry index for the ascending and descending phases separately then the two curves of Fig.4. are obtained. The meaning of these curves is the same as that of Fig.3, as is demonstrated by the arrows connecting both points of each cycle: they have the same directions and lengths as those of Fig.3. The explanation is that if e.g. the northern cycle precedes then the asymmetry index is higher in the ascending phase and lower in the descending phase (as in the cycles 12-15 and 20-23) and the opposite is the case if the southern cycle precedes (cycles 16-19). Figs 2 and 4 demonstrate the same phenomenon with two different methods.

   \begin{figure}
  \begin{center}
   \epsfig{file=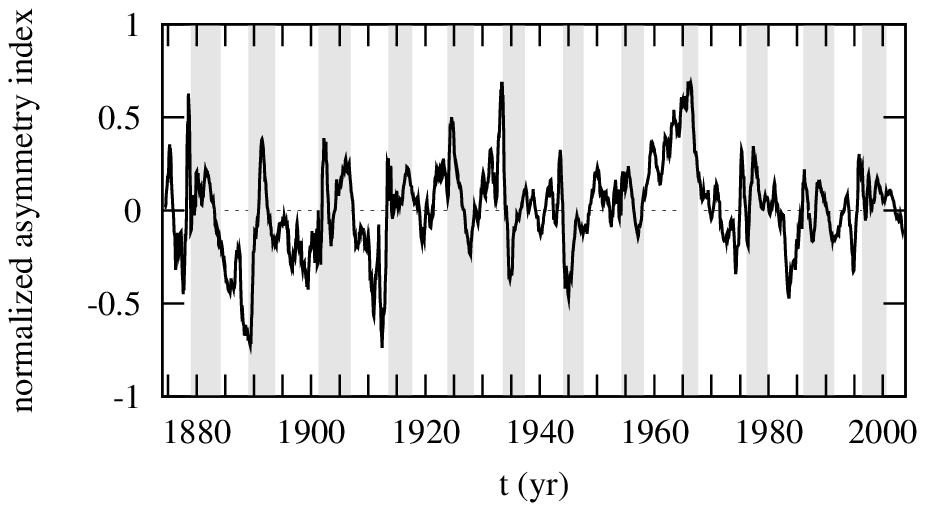,angle=-90,width=11cm}
  \end{center}
  \caption{Monthly values of the asymmetry index between 1873-2003. The ascending and descending phases of the cycles are separated by grey and white stripes.}
\end{figure}

   \begin{figure}
  \begin{center}
   \epsfig{file=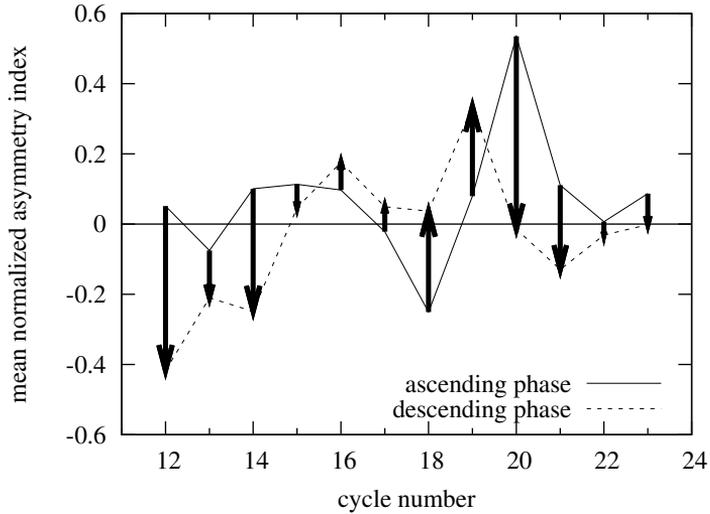,angle=-90,width=10cm}
  \end{center}
  \caption{Variation of the asymmetry index in the ascending and descending phases
of cycles 12-23. The orientations of arrows are identical to those of fig. 2. and their relative lengths are fairly similar to those relative lengths.}
\end{figure} 

The alternating hemispheric precedence, switching after 4 cycles, can be considered as a torsional oscillation of the solar global magnetic field. The term 'torsional oscillation' is primarily used for the fine structure of the differential rotation (Howard and LaBonte, 1980), i.e. for a velocity field but it is also used for magnetic flux ropes, e.g. for those of sunspots (Gopasyuk, 2005) so the proper use of the term is not misleading. 

In the present case the hemispheric initiating role seems to be alternating by 4 cycles. This will deserve a closer scrutiny, here we only mention a possible further direction. The advanced state of a cycle may also be related to the tilt angle of the active region magnetic axes (connecting the leading and following parts). Fig.5 depicts the mean tilt angles for the northern and southern hemispheres in cycles 15-23. The data of the last two cycles are taken from the DPD (Gy\H ori et al, 2009) the rest is taken from the Mount Wilson and Kodaikanal tilt angle datasets (Howard et al, 1984, Sivaraman et al. 1999). Until cycle 19 (period of southern precedence) the northern angle is definitely higher whereas from cycle 20 (period of northern precedence) the values of the two hemispheres progress close to each other. The only apparent exception is cycle 15 which has a small northern precedence and in Fig.5 it seems to belong to the group of cycles 16-19, when southern cycles precede. Anyhow, it may be conjectured that the hemispheric phase differences can be related to further signatures of the cycle progress such as the active region tilt angle or the hemispheric profiles of the differential rotation. These questions need further study.

  \begin{figure}
  \begin{center}
   \epsfig{file=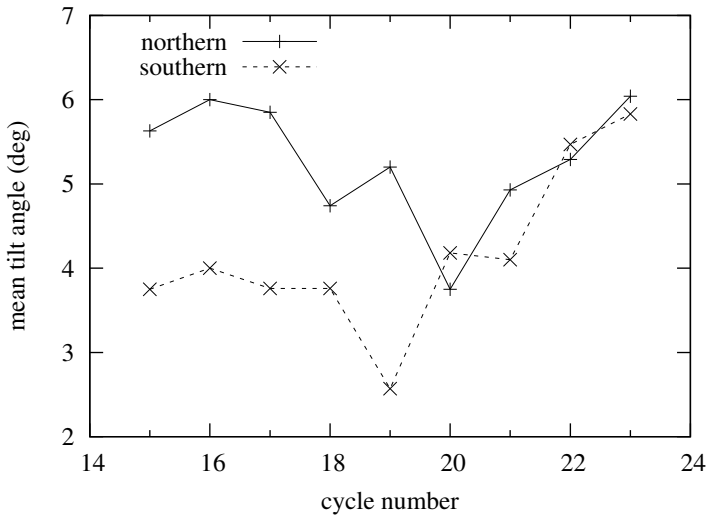,angle=-90,width=10cm}
  \end{center}
  \caption{Mean tilt angles of active region magnetic axes in cycles 15-23 in the northern and southern hemispheres}
\end{figure}

\section{Discussion}

As was mentioned in the introduction several periods were published in earlier reports. Ballester et al. (2005) as well as Zolotova and Ponyavin (2007) published a 43-year peak in the asymmetry periodogram. Their methods are Fourier analysis and Cross-Recurrence Technique and they apparently may have found a harmonic of the period reported here. The rest of the papers report fairly different periods. Vizoso and Ballester (1990) reported 3.27 years, not detected later by Ballester et al (2005), they do not resolve the controversy. Chang (2007) published a period between 9-12 years, similarly to the 11.8 years period obtained by Vizoso and Ballester (1990) which is obviously the signal of the cycle. 

As for the longer periods, Braj\v{s}a et al. (2009) published a period of 70 years. They admit, this is shorter than the Gleissberg cycle and consequently from that published here. The reason of the discrepancy may be that they simply focused on the time behaviour of asymmetry index without considering the individual cycles and their phases. Even longer periods were reported by Forg\'acs-Dajka et al. (2004): longer than 100 years and by Li et al. (2002, 2009): as long as 12 Schwabe cycles. These periods are already too close to the length of the available datasets so it is difficult to interpret them.

Two papers report most similar results to ours. Waldmeier (1971) also published an eight cycle long period of hemispheric phase-lags. It is interesting that he followed a quite different method but obtained an almost identical behaviour to our results. He computed the cyclic average of the yearly mean north-south latitudinal differences of sunspot groups, which is a quite different indicator of the hemispheric precedence than ours, and obtained a very similar time profile. Moreover, his period covers the cycles 10 through 20 and the cycles 10-11 have southern precedence and this corroborates the existence of the 8-cycle period presented here. The other similar work was published by Zolotova and Ponyavin (2009). Their method is also different from ours and that of Waldmeier, e.g., they consider the possibility of phase-lag sign reversal also within the cycle, namely at the maximum of cycle 16. This is the reason why their period is not bound to the length of eight entire cycles.

The presented feature, the alternating hemispheric precedence of the cycle progress over eight Schwabe cycles, raises the question of the long-term variations of solar dynamo. The self-consistent dynamo theories mainly deal with the reconstruction of the observed features of the individual cycles or two consecutive cycles and the central problem is the toroidal-poloidal transit in any types of models. The literature of solar dynamo is enormous, a recent review on the different types of dynamo models is given by Jonas et al. (2009) with ample literature therein. The authors show that all types of dynamo, even the most popular mean field models have serious difficulties in describing the way of building up the poloidal field. The recently very popular forecast attempts encounter an even more difficult problem: some information should be retained by an unspecified solar memory for at least two cycles to be able to esteem the outlooks for the next activity, see e.g. Solanki et al. (2002) and Hathaway (2009). The above presented feature, however, would need a much longer memory which could ensure that after eight cycles the precedence should switch again from northern to southern type. This seems to be fairly difficult to be incorporated into the recent dynamo models. For this reason it may be timely to reconsider the possible role of certain external impacts on the dynamo regime as a modulating effect. As an example, this may also be the case in seeking a plausible explanation for the long-term modulation of the cycle strength, in particular, the Maunder minimum. Following the work of Jose (1965) and Fairbridge and Shirley (1987), it is conspicuous that the inertial motion of the solar center of mass and the solar activity over several centuries have remarkable similarities. Of course, this doesn't mean that the solar inertial motion would be the cause of the solar cycles but its modulating effect cannot be excluded. Similarly, one can conjecture that the present variation could also be resulted by the heliographic latitudinal motion of the solar system barycenter, but this will also be addressed in a further study.

\section*{Acknowledgements} 
The research leading to these results has received funding from the European Community's Seventh Framework Programme (FP7/2007-2013) under grant agreement No. 218816.

\newpage

\section*{References}
\begin{itemize}
\small
\itemsep -2pt
\itemindent -20pt

\item[] Ballester, J. L., Oliver, R., Carbonell, M.: 2005, {\it \aap},  {\bf 431}, L5.
\item[] Berdyugina, S. V., Usoskin, I. G., 2003, {\it \aap}, {\bf 405}, 1121.
\item[] Braj\v{s}a, R., W\"ohl, H., Hanslmeier, A., Verbanac, G., Ru\v{z}djak, D., Cliver, E., Svalgaard, L., Roth, M.: 2009, {\it \aap},  {\bf 496}, 855.
\item[] Chang, H-Y.: 2007, {\it New Ast.},  {\bf 13}, 195. 
\item[] Fairbridge, R. W., Shirley, J. H.: 1987, {\it \solphys}, {\bf 110}, 191.
\item[] Forg\'acs-Dajka, E., Major, B., Borkovits, T.: 2004, {\it \aap},  {\bf 424}, 311.
\item[] Gy\H ori, L., Baranyi, T., Ludm\'any, A. et al., 2009, Debrecen Photoheliographic Data for 1986-2003, see: http://fenyi.solarobs.unideb.hu/DPD/index.html 
\item[] Gopasyuk, S. I., Gopasyuk, O. S.: 2005, {\it \solphys},  {\bf 231}, 11.
\item[] Hathaway, D. H.: 2009, Space Sci. Rev., DOI 10.1007/s11214-008-9430-4
\item[] Howard, R., LaBonte, B. J.: 1980, {\it \apj}, {\bf 239}, L33. 
\item[] Howard, R., Gilman, P., Gilman, P.: 1984, {\it \apj}, {\bf 283}, 373.
\item[] Jones, C. A., Thompson, M. J., Tobias, S. M.: 2009, Space Sci. Rev., DOI 10.1007/s11214-009-9579-5
\item[] Kislovodsk sunspot group reports 1954-2007, \\ see: http://158.250.29.123:8000/web/Soln\_Dann/
\item[] Jose, P. D.: 1965, {\it Astron.J.}, {\bf 70}, 193. 
\item[] Li, K. J., Wang, J. X., Xiong, S. Y., Liang, H. F., Yun, H. S., Gu, X. M.: 2002, {\it \aap},  {\bf 383}, 648.
\item[] Li, K. J., Gao, P. X., Zhan, L. S.: 2009, {\it \solphys},  {\bf 254}, 145.
\item[] Murak\"ozy, J., Ludm\'any, A.: 2008, {\it \aap}, {\bf 486}, 1003.
\item[] Newton, H. W., Milsom, A. S.: 1955, \mnras, 115, 398.
\item[] Ol\'ah, K., Koll\'ath, Z. and 9. co-authors: 2009, {\it \aap}, {\bf 501}, 703.
\item[] Royal Observatory, Greenwich, Greenwich Photoheliographic Results, 1874-1976, in 103 volumes, \\
see: http://www.ngdc.noaa.gov/stp/SOLAR/ftpsunspotregions.html
\item[] Sivaraman, K.R., Gupta, S.S., Howard, R.F.: 1999, {\it \solphys}, {\bf 189}, 69.
see at: http://www.ngdc.noaa.gov/stp/SOLAR/ftpsunspotregions.html
\item[] Solanki, S. K., Krivova, N. A., Sch\"ussler M., Fligge, M.: 2002, {\it \aap}, {\bf 396} 1029.
\item[] Vizoso, G., Ballester, J. L.: 1990, {\it \aap},  {\bf 229}, 540.
\item[] Waldmeier, M.: 1971, \solphys, {\bf 20}, 332. 
\item[] Zolotova, N. V., Ponyavin, D. I.: 2007, {\it \solphys},  {\bf 243}, 193.
\item[] Zolotova, N. V., Ponyavin, D. I., Marwan, N., Kurths, J.: 2009, \aap, {\bf 503}, 197.
\end{itemize}

\end{document}